\documentclass[12pt,preprint]{aastex}
\begin{document}
\title{DISKS AND PLANETS AROUND MASSIVE WHITE DWARFS}
\author{M.~Livio,\altaffilmark{1,2} J.~E.\ Pringle,\altaffilmark{2,3} K.~Wood\/\altaffilmark{2}}

\altaffiltext{1}{Space Telescope Science Institute, 3700 San Martin Drive, Baltimore, MD 21218, USA}
\altaffiltext{2}{Physics and Astronomy, University of St.~Andrews, St.~Andrews KY16 9SS, Scotland, UK}
\altaffiltext{3}{Institute of Astronomy, Madingley Road, Cambridge CB3 0HA, UK}

\begin{abstract}
We predict the existence of dusty disks and possibly CO planets around massive white dwarfs.  We show that the thermal emission from these disks should be detectable in the infrared. The planets may also be detectable either by direct IR imaging, spectroscopy, or using the pulsations of the white dwarfs.
\end{abstract}

\maketitle

\section{INTRODUCTION}
The recent direct imaging of the companion to the brown dwarf 2M1207A, and the estimates of its mass (2--5 Jupiter masses) open a new era in extrasolar planet research (Schneider et~al.\ 2005; Chauvin et~al.\ 2005).  At the same time, \emph{Spitzer} observations are now revealing detailed properties of dusty disks around young stellar objects (e.g., Meyer et~al.\ 2004).

In the present letter we propose that disks and planets may be found in an unexpected place---around massive white dwarfs.  Our model is presented in Section~2 and some observational tests are discussed in Section~3.

\section{MERGING WHITE DWARFS, PLANETS AND DUST DISKS}

The merger of two white dwarfs is an inevitable outcome of binary star evolution in some binary systems (see, e.g., Iben \& Livio 1993 for a review).  Liebert, Bergeron, and Holberg (2005) examined a sample of 348 DA white dwarfs (WDs). They found that the mass distribution has three components: (i)~A peak centered around 0.6~$M_{\odot}$, (ii)~a low-mass component around 0.4~$M_{\odot}$, and (iii)~a high-mass component extending above 0.8~$M_{\odot}$.  The high-mass component was estimated to contribute about 
15\% of the WD formation rate. Liebert et~al.\ (2005) also concluded that most ($\gtrsim80$\%) of the high-mass objects formed via mergers of two lower-mass WDs. The merger process itself has been studied in some detail (e.g., Benz et~al.\ 1990; Mochkovitch \& Livio 1990). It leads to the total dissipation of the lighter of the two WDs (which fills its Roche lobe first) within a few orbital periods. The dissipated WD material forms a disk around the more massive component, most of which is eventually accreted onto the heavier member to form the massive WDs found by Liebert et~al. However, in order for this disk material to be accreted at all, some fraction of the matter in the disk needs to take up the angular momentum. The matter into which the angular momentum is deposited spreads out to form a disk that is much lower in mass, but much larger in radius.  Note, however, that the remaining disk is essentially 100\% metals, since it typically represents the composition of a CO WD.

To model this process we start with two identical white dwarfs, each of mass $M_{WD} = 0.5~M_\odot$ in a circular orbit around each other, and filling their Roche lobes. These merge to give a new white dwarf whose mass is essentially equal to 
$M_{\rm NWD} = 2~M_{WD} = 1~M_\odot$. This implies that if the semi-major axis of the binary is $a$, the mean white dwarf radii are $R_{\rm WD} = 0.38 a$. In this case the orbital angular momentum is
\begin{equation}
J_{\rm orb} = \frac{1}{4} M_{\rm NWD} (G M_{\rm NWD} a)^{1/2}~~.
\end{equation}
It is straightforward to show that this angular momentum cannot be taken into the spin of the newly formed white dwarf (since the breakup speed would be exceeded by a factor of around 3). We therefore model the ensuing dynamics of the merger as an accretion disk with fixed angular momentum $J_{\rm orb}$ around a central point mass $M_{\rm NWD}$. The time evolution of such disks is well known (e.g., Pringle 1981; Pringle 1991). The exact behavior depends on the details of the dissipative processes within the disk, but at late times the outer radius of the disk expands to take up the angular momentum, and the inner regions behave as a steady accretion disk, but with gradually decreasing accretion rate. In this sense, the disk dynamics can be expected to be akin to that of the usual protostellar disks, in which all the accretion comes from large radii, but at a steadily declining rate. For simplicity we assume that the surface density profile of the disk is of the form
\begin{equation}
\Sigma \propto R^{-\alpha}~~,
\end{equation}
where we require $\alpha < 2$. For example, the usual solar nebula profile is assumed to have $\alpha \approx 3/2$ (Armitage et~al.\ 2002), whereas if we adopt the Lin \& Papaloizou (1985) prescription for the viscosity in cool protostellar disks of $\nu\propto\Sigma^2$ we would obtain $\alpha \approx 0$. With this profile, angular momentum of the disk is given in terms of the disk radius $R_d$ and the disk mass $M_d$ by
\begin{equation}
J_d = \xi M_d (G M_{NWD} R_d)^{1/2}~~,
\end{equation}
where $\xi = (2-\alpha)/(\frac{5}{2} -\alpha)$. We shall adopt $\alpha = 3/2$ and $\xi = 1/2$ as typical parameters.

If we equate the angular momentum of the disk to that of the original binary we find that the disk mass is given in terms of the disk radius by 
\begin{equation}
M_d = 0.81 \left( \frac{\xi}{1/2} \right)^{-1} M_{\rm NWD} 
\left(\frac{R_{\rm WD}}{R_d} \right)^{1/2}~~.
\end{equation}

We now need to ask at what stage planets might start to form in such a disk. We have already argued that a disk of this sort does not differ in its dynamics greatly from a standard protostellar disk. If this were a disk with standard cosmic abundances then it might seem
reasonable to assume that planetesimal formation sets in when the disk radius exceeds the snow-line, which is the radius outside which the disk temperature equals the ice condensation temperature of $T_c \approx 170$~K. But this disk does not have solar composition, and in particular lacks the hydrogen to produce the water for the `snow.' In this situation it seems sensible to adopt the dust grain condensation temperature of $T_c \approx 1600$~K as the equivalent of the snow line. Adopting the temperature profile for an optically thin disk given by
\begin{equation}
T_d = 2.8 \times 10^2 \left( \frac{R}{1~\mathrm{AU}} \right)^{-1/2}
\left(
\frac{L}{L_\odot} \right)^{1/4}~\mathrm{K}~~,
\end{equation}
gives the condensation radius as
\begin{equation}
R_\mathrm{dust} = 0.0306 \left( \frac{L}{L_\odot} \right)^{1/2} \left(
\frac{T_c}{1600~\mathrm{K}} \right)^{-2}~\mathrm{AU}~~.
\end{equation}

Taking the radius of the newly formed white dwarf to be $R_{\rm NWD} = 6 \times 10^8$~cm, and its temperature to be $T_{\rm NWD} = 50,\!000$~K, we find that the dust condensation line occurs at radius
\begin{equation}
R_\mathrm{dust} = 0.02 \left( \frac{R_\mathrm{NWD}}{6\times10^8~\mathrm{cm}} \right)
\left( \frac{T_\mathrm{NWD}}{50,000~\mathrm{K}} \right)^2 \left(
\frac{T_c}{1600~\mathrm{K}} \right)^{-2}~\mathrm{AU}~~.
\end{equation}

When this radius is reached, the mass in the disk is
\begin{equation}
M_d = 0.047 \left( \frac{\xi}{1/2} \right)^{-1} \frac{M_\mathrm{NWD}}{M_\odot}
\left( \frac{0.6 R_\mathrm{WD}}{ R_\mathrm{NWD}} \right)^{1/2} 
\left( \frac{T_\mathrm{NWD}}{50,000~\mathrm{K}} \right)^{-1} 
\left( \frac{T_c}{1600~\mathrm{K}} \right)^{-1}~M_\odot~~.
\end{equation}
Since the mass is concentrated at large radii, we may take this mass as an estimate of the mass available for forming planets.  As the radius expands to $\sim$1~AU, the mass in the disk is $\sim\!0.007~M_{\odot}$.

In a recent, important work, Fischer and Valenti (2005) have shown that the probability to host a planet increases quite dramatically
with the metallicity of the host star (approximately as [Fe/H]$^{1.8}$, where [Fe/H] represents the iron abundance). If this result can be extrapolated to the metallicities under discussion here, it implies that planets formation would be highly efficient in disks of the kind we are considering. 

The only other situation we are aware of which is analogous to the planet formation picture we have described above is the formation of planets around the pulsar PSR1257+12 (Wolszcan \& Frail 1992). The planets in this system, presumably around a neutron star of mass around $M_\mathrm{NS} \approx 1.4~M_\odot$, move in almost circular orbits, and so most likely formed in a disk. They have masses of 2.8~$M_\mathrm{Earth}$ at 0.47~AU and 3.4~$M_\mathrm{Earth}$ at 0.36~AU. The disk in which the planets have formed is thought to have come from the disruption of a low mass companion (Stevens, Rees \& Podsiadlowski 1992) of mass $M_2 \approx
0.016~M_\odot$ (King et~al.\ 2005). In this case, we calculate that the fraction of the original orbital angular momentum which ended up in these two planets is $f \sim 1$\%. If the planets, being Earth-mass, are also Earth-like, then this implies a high efficiency of converting
the initial angular momentum stored in elements heavier that H and He into planets.

\section{OBSERVATIONAL CONSEQUENCES}

One of the predictions of the proposed scenario is the potential existence of dusty disks around massive WDs. Our calculations suggest that a typical circumstellar dust disk will have a mass and radius of $M_d\sim 0.007~M_\odot$ and $R_d\sim 1$~AU, respectively.  The disk will be gas poor and the dust will reprocess radiation from the white dwarf to produce an infrared excess spectral energy distribution (SED).  Figure~1 shows a simulated SED for a dust disk around a white dwarf (assumed to be at a distance of 12~pc, typical of nearby WDs; Holberg, Oswalt \& Sion 2002), demonstrating that the thermal emission from the disk will be detectable above the white dwarf photosphere at infrared wavelengths (the different plots are evenly spaced in the cosine of the inclination).  In this simulation, the white dwarf (after merger) has $M_\mathrm{NWD}=M_\odot$, $T_\mathrm{NWD}=5\times10^4$~K, $R_\mathrm{NWD}=6\times10^8$~cm, and a Kurucz model atmosphere was used. Using a $5\times10^4$~K black body did not result in any significant changes. The disk has a dust mass $M_d = 0.007~M_\odot$, outer radius $R_d = 1$~AU, and inner radius corresponding to a dust destruction temperature of 1600~K.  The dust opacity and scattering properties are assumed to be those for interstellar dust (e.g., Kim, Martin, \& Henry 1994), which gives the prominent $10~\mu$m silicate feature in the simulated SED.  If the dust in the disk has grown to sizes larger than typical ISM grains, the opacity will be grayer and the silicate features less prominent (e.g., Wood et~al.\ 2002). We assume the dust is in vertical hydrostatic equilibrium and the disk has a radial surface density $\Sigma\sim r^{-3/2}$.  The disk temperature, vertical hydrostatic density structure, and emergent SED are computed using the Monte Carlo radiation transfer code described in Walker et~al.\ (2004).  If the dust is not in vertical hydrostatic equilibrium and has settled to the midplane, the  infrared excess emission will be less than shown in Figure~1, but will still be detectable above the photosphere and resemble that from a flat reprocessing disk (e.g., Adams, Lada, \& Shu 1987).  Note that massive WDs are considerably smaller than their low-mass counterparts, and hence, at a given temperature they are less luminous.

Disks that are extremely metal-rich may also be expected to form planets. Following the pioneering work of Becklin \& Zuckerman (1988), a few more extensive searches for planets around WDs have already started.  The search methods include IR imaging, IR spectroscopy, and the use of WD pulsations. Clarke \& Burleigh (2004; see also Burleigh, Clarke \& Hodgkin 2002) describe the first results from a deep infrared imaging ($J\sim24$) campaign of twenty-four WDs. They found two objects that appear to have the same proper motion as the WDs (at least at the 1$\sigma$ level), and whose luminosities are consistent with masses in the range 7--10~M$_\mathrm{Jup}$. The Clarke \& Burleigh search aims at planets that existed around the WD progenitor, and which have survived stellar evolution. The current paper identifies \emph{massive WDs} as high-probability hosts of dusty disks or planets, with these disks and planets having formed during a late evolutionary phase, and having unusual compositions.

A second existing search for planets around WDs uses the pulsations of the WDs as intrinsic clocks (Mullally et~al.\ 2003). The idea here is that hot, DAV-type WDs near 12,000~K pulsate with an extraordinarily stable pulsation period. The reflex orbital motion causes a rate of change $\dot P$ in the period $P$ of order (e.g., Kepler et~al.\ 1991)
\begin{equation}
\frac{\dot P}{P}=\frac{G}{c} \frac{PM_p \sin i}{r^2_p}~~,
\end{equation}
where $G$ and $c$ are the gravitational constant and speed of light, respectively, and $i$ is the orbital inclination. Again, in the present paper we have identified massive WDs as promising targets. While WD cooling also produces a period change, this change is expected to be monotonic, while the change caused by a planet is periodic.

Finally, Dobbie et~al.\ (2005) have used a near-infrared spectroscopic analysis of eight white dwarfs, in an attempt to search for low-mass companions. They were able to place upper limits on putative companions at substellar masses ($\sim$0.07~$M_{\odot}$). We have also examined the WDs with cool companions reported by Wachter et~al.\ (2003), who used 2MASS photometry.  However, none of the confirmed high-mass WDs appears to be on that list.

We should also note that the probability of even a Jovian-mass planet (most likely of CO composition) at 0.01~AU eclipsing its host WD is rather small, $\sim$2\%. Note, however, that a Jovian planet of CO composition will be larger (by more than a factor 6) than a 1.2~$M_{\odot}$ WD (e.g., Zapolsky \& Salpeter 1969), thus enabling it to produce a total eclipse.

If future observations fail to detect any disks or planets around massive WDs, this may mean that the formation of dust is somehow suppressed in these somewhat unusual, hydrogen-poor environments.

\begin{acknowledgments}
We are grateful to Jim Liebert and Keith Horne for helpful discussions. M.L.\ thanks the Dept.\ of Astronomy at the University of St.\  Andrews for its hospitality during a Carnegie Centenary Professorship.
\end{acknowledgments}

\begin{figure}
\includegraphics[height=\textwidth,angle=90]{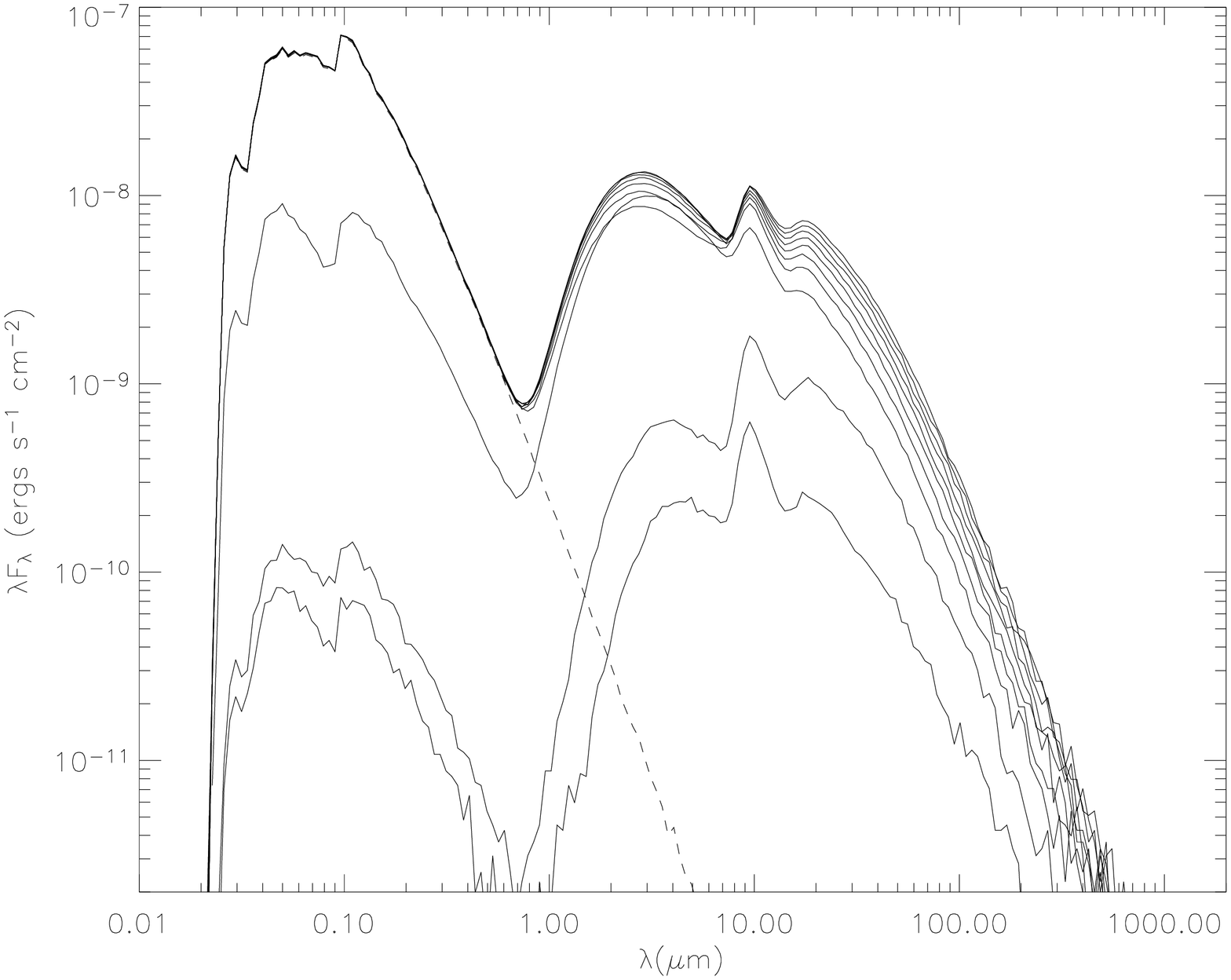}
\caption{
Simulated spectral energy distribution for a 1~AU radius dust disk surrounding a white dwarf.  The dashed line is the input stellar 
spectrum and the solid lines show the characteristic infrared excess emission for ten viewing angles evenly spaced in $\cos i$ (pole-on viewing is the uppermost solid line). At high inclinations, the white dwarf is occulted by the disk and the optical spectrum is all scattered light.}
\end{figure}

\end{document}